\documentclass[letterpaper,11pt,fleqn]{article}

\usepackage[utf8]{inputenc} 
\usepackage{fullpage,physics,graphicx,xcolor,bm,bbm,authblk, amsmath,amsfonts,amsbsy,amssymb,hyperref}
\usepackage[numbers, square, sort&compress]{natbib}
\hypersetup{colorlinks = true, linkbordercolor = white, linkcolor = blue, citecolor = blue}
\newcommand{\equa}[1]{Eq.~(\ref{#1})} \newcommand{\equas}[1]{Eqs.~(\ref{#1})}
\newcommand{\equass}[2]{Eqs.~(\ref{#1})-(\ref{#2})}
\newcommand{\equasa}[2]{Eqs.~(\ref{#1}) and (\ref{#2})}
\newcommand{\Pe}{{\rm Pe}} \newcommand{\Rey}{{\rm Re}} \newcommand{\Ge}{{\rm Ge}} \newcommand{\RR}{{\cal R}}
\newcommand{\eqn}[2]{\begin{gather}
#1
\label{#2}
\end{gather}
}
\newcommand{\gat}[2]{\begin{subequations}\label{#2}\begin{gather}
#1
\end{gather}\end{subequations}
}
\title{\bf Viscous heating and instability of the adiabatic\\ buoyant flows in a horizontal channel}
\author{\bf A. Barletta\footnote{Corresponding author: \texttt{antonio.barletta@unibo.it}}\ ; M. Celli}
\affil{\small Department of Industrial Engineering, Alma Mater Studiorum Universit\`a di Bologna,\\
Viale Risorgimento 2, 40136 Bologna, Italy}
\author{\bf D. A. S. Rees}
\affil{\small Department of Mechanical Engineering, University of Bath, Bath BA2 7AY, U.K.}
\date{\small\today} % Activate to display a given date or no date (if empty),
\begin{document}

\maketitle

\begin{abstract}
\noindent The stability of buoyant flows occurring in the mixed convection regime for a viscous fluid in a horizontal plane-parallel channel with adiabatic walls is investigated. The basic flow features a parallel velocity field under stationary state conditions. There exists a duality of flows, for every prescribed value of the mass flow rate across the channel cross-section, caused by the combined actions of viscous dissipation and of the buoyancy force. As pointed out in a previous study, only the primary branch of the dual solutions is compatible with the Oberbeck-Boussinesq approximation. Thus, the stability analysis will be focussed on the stability of such flows. The onset of the thermal instability with small-amplitude perturbations of the basic flow is investigated by assuming a very large Prandtl number, which is equivalent to a creeping flow regime. The neutral stability curves and the critical parametric conditions for the onset of instability are determined numerically.\\[0.5cm]
\end{abstract}

\newpage

\section{Introduction}
The instability of stationary and fully-developed flows in a plane channel is a classical topic discussed in a plethora of textbooks on fluid mechanics. Among the many, we mention \citet{drazin2004hydrodynamic} and  \citet{kundu2016fluid}. The instability arises as hydrodynamic in nature when there is no thermal forcing on the flow via the boundary conditions, with the critical conditions for linear instability being determined through the solution of the Orr-Sommerfeld eigenvalue problem \cite{kundu2016fluid, drazin2004hydrodynamic}. A different conclusion is drawn when, even in the absence of an external temperature difference between the plane boundary walls, the frictional heating associated with the viscous flow is taken into account. In this case, the thermally-induced instability acts via a temperature coupling term in the local momentum balance equation. Such a term can be the viscous force as the fluid viscosity is temperature-dependent \cite{joseph1964variable, joseph1965stability}. An alternative scenario is the buoyancy effect, where the temperature coupling term in the momentum balance is the gravity force with a temperature-dependent fluid density modelled according to the Oberbeck-Boussinesq approximation \cite{barletta2010convection, barletta2011onset, barletta2015thermal, miklavvcivc2015stability, barletta2016fully, lund2019effect, requile2020weakly, lund2021magnetohydrodynamic, ReisAlves2021, barletta2021energy, ali2022soret}.

The combined effects of buoyancy and viscous heating may yield situations where the stationary flows in a channel or duct caused by a given dynamic input, either a prescribed pressure gradient or a prescribed mass flow rate, are dual. The duality is usually accompanied by a merging between the solution branches which produces a maximum parametric condition in terms of either pressure gradient or mass flow rate above which no stationary flow is possible. This behaviour is widely documented in the literature \cite{turcotte1982multiple, BARLETTA20054835, miklavvcivc2015stability, barletta2016fully, lund2019effect, lund2021magnetohydrodynamic, barletta2022mixed}. We mention that dual or, more generally, multiple solutions are a consequence of the nonlinearity of buoyant flows and their existence does not necessarily imply the inclusion of the viscous dissipation term in the local energy balance equation \cite{wilks_bramley_1981, naganthran2020dual}.

The aim of this paper is the stability analysis of the primary branch of dual solutions found by \citet{barletta2022mixed}. In fact, these authors pointed out that the secondary branch refers to conditions hardly compatible with the Oberbeck-Boussinesq approximation scheme underlying their determination. The flow conditions employed in this analysis involve the fully-developed regime in a horizontal channel with thermally insulated walls. The interplay between the buoyancy force and the viscous heating of the channel flow leads to a mixed convection scenario where the velocity displays a departure from the Poiseuille profile, with a duality of solutions for every prescribed mass flow rate across the channel. Though widely described in \citet{barletta2022mixed}, the main features of the dual flows are surveyed also in this paper for self-containedness of our presentation and for a precise definition of which basic flow we assume when testing the transition to instability. The character of our stability analysis reflects a similar study published some years ago and relative to Darcy's flow in a fluid-saturated porous material \cite{barletta2009stability}. As we are interested in the destabilising action of the viscous dissipation term in the energy balance, we will assume high viscosity and low diffusivity properties of the fluid, which means a very large Prandtl number. This assumption simplifies the governing equations for the perturbations of the basic solution as they are formulated for a creeping flow scheme with the inertial term in the momentum balance turning out to be negligible. A similar approach was followed also in \citet{barletta2010convection} and in \citet{barletta2011onset}. The focus on the creeping flow regime makes our analysis completely different from the hydrodyncombinedamic stability analysis of the Poiseuille flow as based on the Orr-Sommerfeld eigenvalue problem. In fact, the hydrodynamic instability is relative to a condition where the inertial term of the momentum balance becomes utterly important for the emergence of the flow instability so that, in that case, the creeping flow scheme is inadequate \cite{drazin2004hydrodynamic, kundu2016fluid}. 

An important governing parameter in the forthcoming analysis is the Gebhart number, $\Ge$. This parameter is often employed in the literature where buoyant flows are studied by including the effect of viscous dissipation, though several authors prefer calling this parameter dissipation number, $\rm Di$. Indeed, the definitions of $\Ge$ and $\rm Di$ are the same. The former symbol is a recognition of the pioneering study by \citet{gebhart1962}, while the latter is still widely employed in the literature on geophysical flows. Among the many studies regarding the geophysical applications of the natural convection heat transfer with viscous heating, we mention the interesting papers by \citet{kincaid1996role} and by \citet{van1997role}. The paper by \citet{kincaid1996role} provides a model where the excess heat generated in the upper part of Earth mantle during orogenesis is attributed to a viscous dissipation contribution, envisaging also cases where $\rm Di$ is as large as $6$. In \citet{van1997role} the assumption $\rm Di = 0.7$ is made on studying buoyant flows in the mantle by including the effect of viscous dissipation coupled also with the internal heating due to radioactivity and the adiabatic work induced by compression/decompression processes. 

\section{Mathematical Model}

Let us consider a Newtonian fluid flowing in a plane-parallel channel bounded by walls at $z = 0$ and $z = H$. The horizontal $x$ and $y$ directions are unbounded, while the uniform gravitational acceleration is given by $\vb{g} = - g\, \vu{e}_z$, where $g$ is the modulus of $\vb{g}$ and $\vu{e}_z$ is the unit vector of the $z$ axis. The boundary walls are both rigid and with a perfect thermal insulation.

\subsection{Governing Equations}
The Oberbeck-Boussinesq approximation can be employed so that the governing equations are written as,
\gat{
\div{\vb{u}} = 0, \label{1a}\\
\pdv{\vb{u}}{t} + (\vb{u} \cdot \grad{)\,\vb{u}} = - \frac{1}{\rho}\, \grad{p} + g \beta \qty(T - T_0)\, \vu{e}_z + \nu\, \nabla^2 \vb{u},\label{1b}\\
\pdv{T}{t} + (\vb{u} \cdot \grad{)\,T} = \alpha \, \nabla^2 T + \frac{\nu}{c}\, \Phi , \label{1c}
}{1}
where $\rho$, $\beta$, $\nu$, $\alpha$ and $c$ are the fluid density, thermal expansion coefficient, kinematic viscosity, thermal diffusivity and specific heat evaluated at the constant reference temperature $T_0$. In \equas{1}, $\vb{u}$ is the velocity, $p$ is the local difference between the pressure and the hydrostatic pressure, $T$ is the temperature and $t$ is the time. Hereafter, $p$ is called the pressure field for brevity. The symbol $\Phi$ denotes the dissipation function which, according to Einstein's notation for the implicit sums over repeated indices, can be expressed as
\eqn{
\Phi = \frac{1}{2} \, \gamma_{ij} \gamma_{ij} \qusing \gamma_{ij} = \pdv{u_i}{x_j} + \pdv{u_j}{x_i} .
}{2}
In \equa{2}, $\gamma_{ij}$ is the shear rate tensor, while $u_i$ and $x_i$ denote the $i$th Cartesian components of the velocity vector $\vb{u} = (u, v, w)$ and of the position vector $\vb{x} = (x,y,z)$.

The boundary conditions imposed at $z=0, H$ express impermeability, no-slip and adiabaticity,
\eqn{
\vb{u} = 0 \qc \pdv{T}{z} = 0 \qfor z=0, H.
}{3}

\subsection{Dimensionless Formulation}
Dimensionless quantities can be defined through the scaling
\eqn{
\frac{\vb{x}}{H} \to \vb{x} \qc \frac{t}{H^2/\alpha} \to t \qc \frac{\vb{u}}{\alpha/H} \to \vb{u} \qc \frac{p}{\rho \alpha \nu/H^2} \to p \nonumber\\ 
\frac{T - T_0}{\Delta T} \to T \qc \frac{\Phi}{\alpha^2/H^4} \to \Phi \qusing \Delta T = \frac{\alpha \nu}{g \beta H^3} .
}{4}
On account of \equa{4}, \equasa{1}{3} can be rewritten in a dimensionless form as
\gat{
\div{\vb{u}} = 0, \label{5a}\\
\frac{1}{\Pr} \left[ \pdv{\vb{u}}{t} + (\vb{u} \cdot \grad{)\,\vb{u}} \right] = - \grad{p} + T\, \vu{e}_z + \nabla^2 \vb{u},\label{5b}\\
\pdv{T}{t} + (\vb{u} \cdot \grad{)\,T} = \nabla^2 T + \Ge\, \Phi , \label{5c}
}{5}
with
\eqn{
\vb{u} = 0 \qc \pdv{T}{z} = 0 \qfor z = 0, 1.
}{6}
In \equasa{5b}{5c}, the Prandtl number, $\Pr$, and the Gebhart number, $\Ge$, are defined as
\eqn{
\Pr = \frac{\nu}{\alpha} \qc \Ge = \frac{g \beta H}{c} .
}{7}

\begin{figure}[t]
\centering
\includegraphics[width=\textwidth]{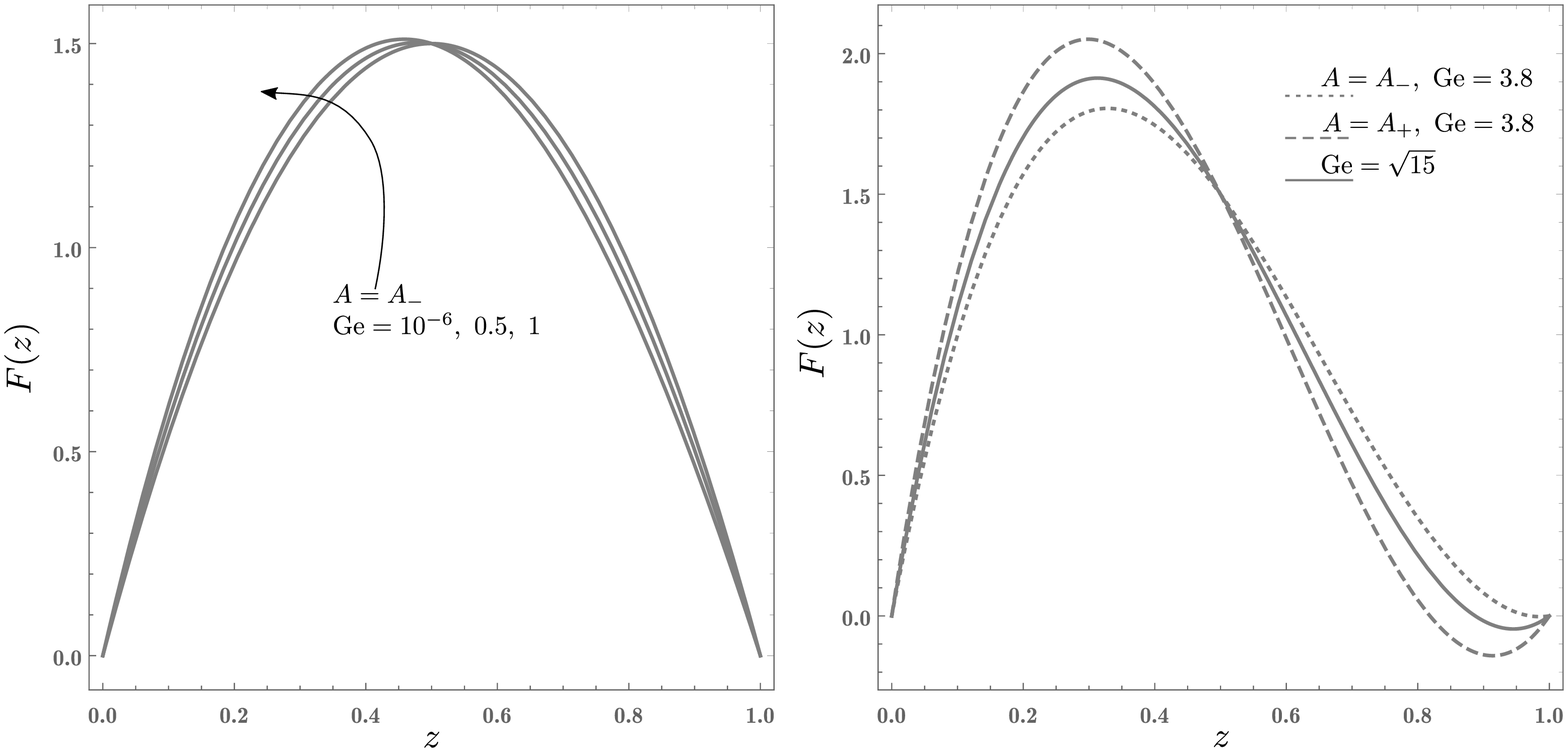}
\caption{\label{fig1}Plots of $F(z)$ for either $A = A_{-}$ or $A = A_{+}$ with different values of $\Ge$}
\end{figure}

\begin{figure}[h!]
\centering
\includegraphics[width=\textwidth]{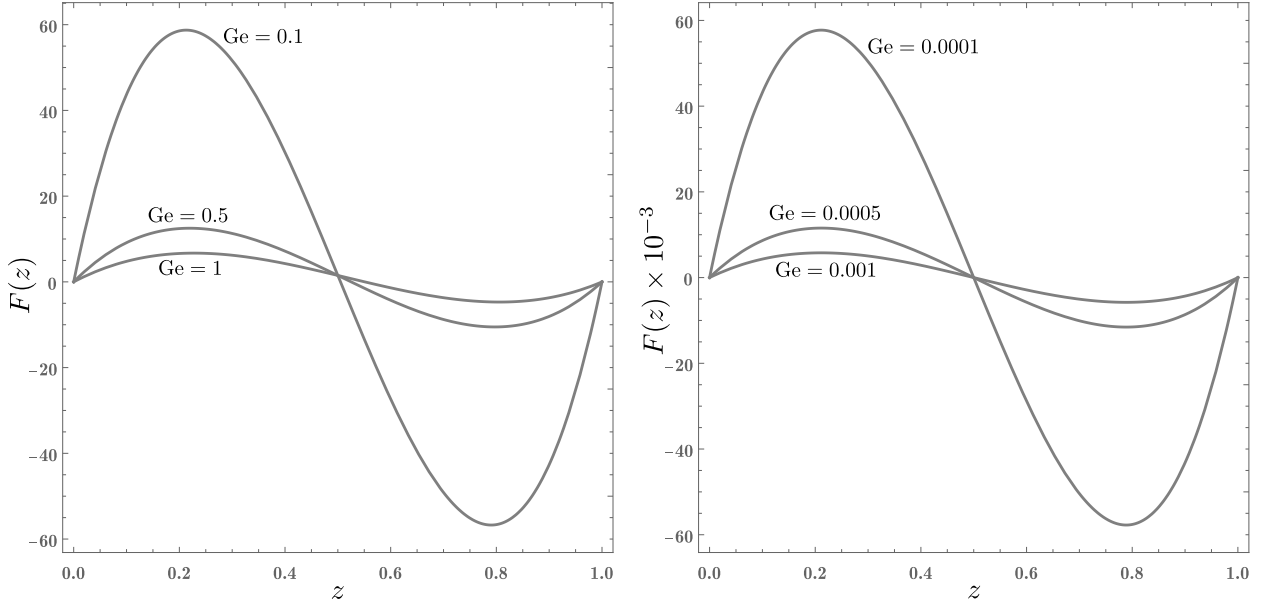}
\caption{\label{fig2}Plots of $F(z)$ for $A = A_{+}$ with different values of $\Ge$}
\end{figure}

\section{Basic Dual Flows}\label{badufl}
In this section, we survey the main features of the dual adiabatic flows relying on the results conveyed in a previous study \cite{barletta2022mixed}.
Stationary solutions of \equasa{5}{6} with a parallel velocity field can be expressed as
\eqn{
u_b = \Pe\, F(z)\, \cos\varphi \qc v_b = \Pe\, F(z)\, \sin\varphi \qc w_b = 0, \nonumber\\
%p_b = \Pe\, F''(z) \left( x \, \cos\varphi + y\, \sin\varphi \right) + G(z) + {\rm constant}, \nonumber\\
T_b = \Pe\, F'''(z) \left( x \, \cos\varphi + y\, \sin\varphi \right) + \Pe^2\, G(z), \nonumber\\
\grad{p_b} = \Big( \Pe\, F''(z)\, \cos\varphi ,\ \Pe\, F''(z)\, \sin\varphi ,\
T_b \Big) ,
}{8}
where the subscript $b$ indicates ``basic solution'', $\Pe$ is the P\'eclet number and the primes denote the derivatives with respect to $z$, while functions $F(z)$ and $G(z)$ are polynomials given by
\eqn{
F(z) = z\, \qty(1 - z)\, \qty[A - 2\, \qty(A - 6)\, z], \nonumber\\
G(z) = \frac{z^2}{10}\, \big\{ - 5\, A^2\, \Ge + 20\, A\, \qty[\qty(A - 4)\, \Ge + A - 6] \, z \nonumber\\
\hspace{3cm}-\, 10 \left[A^2\, (4 \, \Ge + 3) - 30\, A\, \qty(\Ge + 1) + 48\, \Ge + 72 \right] z^2 \nonumber\\
\hspace{4cm}+\, 12\, \qty(A - 6)\, \qty[3 \, \qty(A - 4)\, \Ge + A - 6] \, z^3
- 12 \, \qty( A - 6 )^2\, \Ge\, z^4 \big\}.
}{9}
The constant $A$ can be equal either to $A_{-}$ or $A_{+}$, where
\eqn{
A_{-} = 2\, \frac{3\, \Ge + 15 - \sqrt{15\, \qty(15 - \Ge^2)}}{\Ge} \qc
A_{+} = 2\, \frac{3\, \Ge + 15 + \sqrt{15\, \qty(15 - \Ge^2)}}{\Ge} .
}{10}
Equations~(\ref{8})-(\ref{10}) describe horizontal parallel flows in the $xy$ plane where the velocity field is inclined an angle $\varphi$ to the $x$ axis. The twofold expression of $A = A_{\pm}$ means that there are dual flows corresponding to the same prescribed P\'eclet and Gebhart numbers. Such dual flows exist if $\Ge \le \sqrt{15} \approx 3.87298$ and they coincide only when $\Ge = \sqrt{15}$. We note that such an upper bound for $\Ge$ is an extremely large value in practical cases \cite{barletta2022mixed}. The P\'eclet number is defined in terms of dimensional quantities as
\eqn{
\Pe = \frac{U_0 H}{\alpha} ,
}{11}
where the reference dimensional velocity $U_0$ is intended as the basic average velocity in the horizontal flow direction defined by the unit vector $\qty(\cos\varphi, \sin\varphi, 0)$. In fact, \equa{9} yields 
\eqn{
\int\limits_0^1 F(z) \, \dd z = 1 ,
}{12}
which means that the average dimensionless velocity in the flow direction is equal to $\Pe$.

\begin{figure}[t]
\centering
\includegraphics[width=\textwidth]{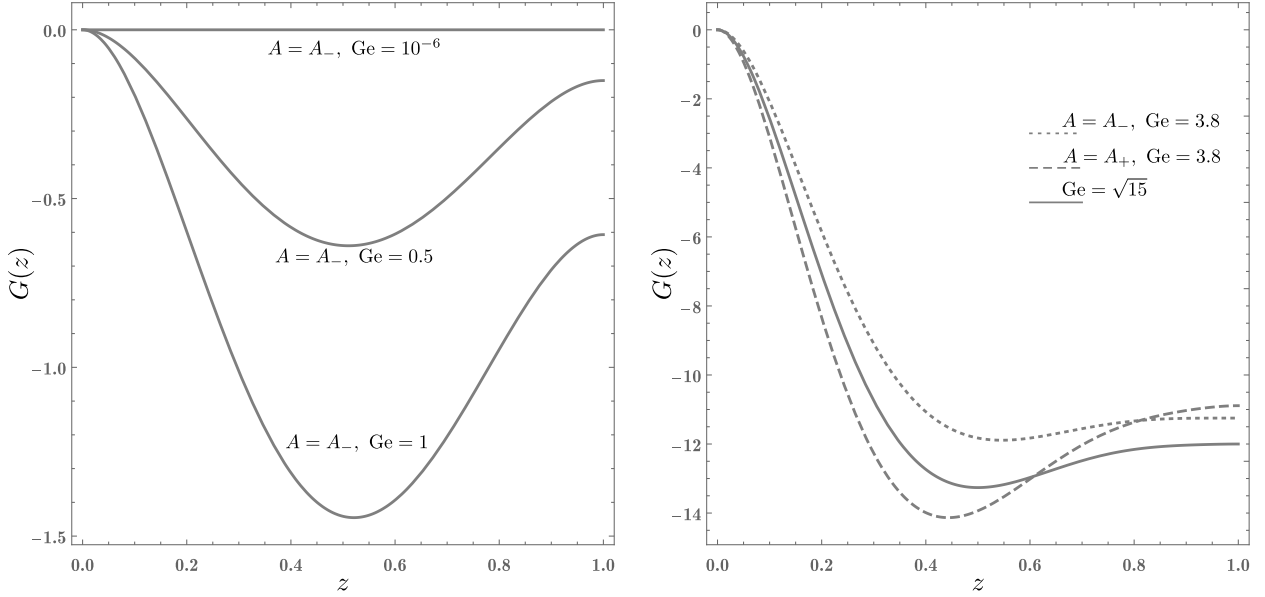}
\caption{\label{fig3}Plots of $G(z)$ for either $A = A_{-}$ or $A = A_{+}$ with different values of $\Ge$}
\end{figure}

\begin{figure}[h!]
\centering
\includegraphics[width=\textwidth]{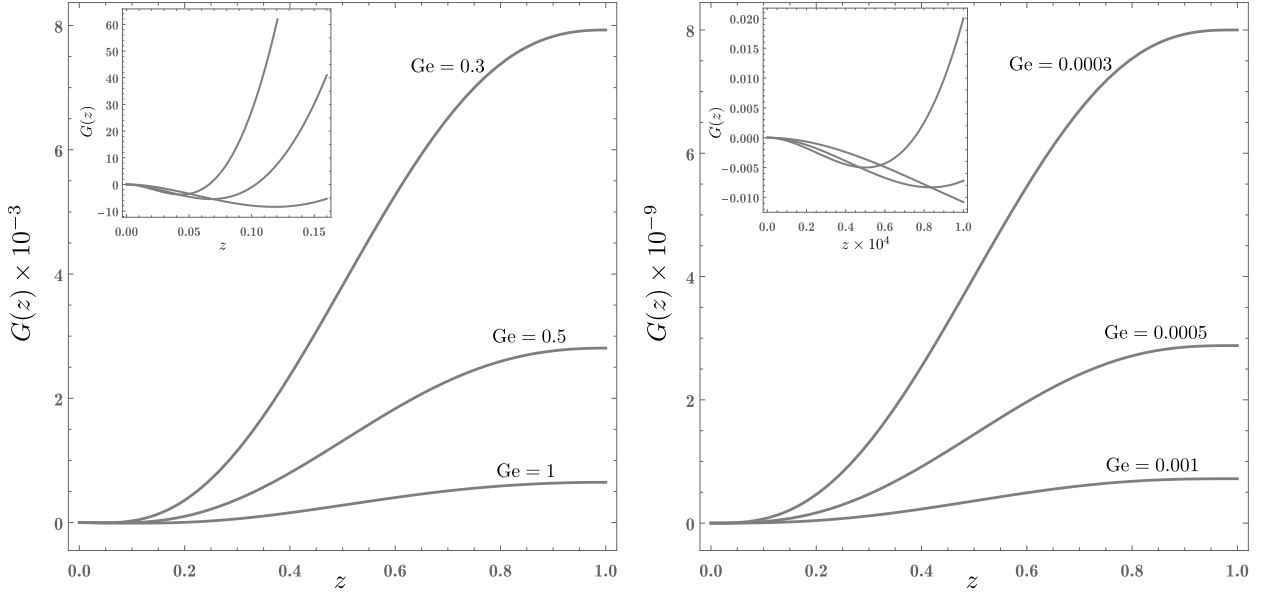}
\caption{\label{fig4}Plots of $G(z)$ for $A = A_{+}$ with different values of $\Ge$}
\end{figure}

It is to be mentioned that function $G(z)$ is defined only up to an arbitrary additive constant which, in \equa{9}, has been fixed so that $G(0)=0$. Such a feature is due to the Neumann boundary conditions for the temperature which leave this constant undetermined. The temperature appears on the right hand side of \equa{5b} next to the $\grad p$ term. Thus, changing the temperature by an additive constant has the physical meaning of changing the reference temperature value employed for the formulation of the Oberbeck-Boussinesq approximation. This change leads to a redetermination of the hydrostatic pressure and to a resulting modification in the dimensional field $p$ which, by definition, is the difference between the pressure and the hydrostatic pressure. In fact, the hydrostatic pressure is $-\, \rho\, g\, z$, where $\rho$ is the fluid density evaluated at the reference temperature $T_0$. It becomes clear that altering the arbitrary additive constant in $G(z)$ influences the basic solution only by modifying, through an overall additive constant, the local values of $T_b$ and, as a consequence of \equa{9}, also of $\partial p_b/\partial z$. On the other hand, the discussion of the stability of the dual basic flows, to be carried out later on, is not affected in any way by the choice of the arbitrary additive constant in the expression of $G(z)$. In fact, it will be shown that the dynamics of perturbations is governed by partial differential equations where $\grad p_b$ is absent and $T_b$ is present only through its gradient. 

The physical effects underlying the basic dual solutions are the imposed horizontal flow rate causing the viscous heating and the buoyancy force induced by the resulting temperature gradient. Such effects become inactive when $\Ge \to 0$, as highlighted by \equa{5c}, and the basic solution becomes an isothermal Poiseuille flow. This conclusion is made evident by taking the power series expansions of $A_{+}$ and $A_{-}$ at small $\Ge$,
\eqn{
A_{-} = 6 + \Ge + \ordersymbol\qty(\Ge^2) \qc
A_{+} = \frac{60}{\Ge} + 6 - \Ge + \ordersymbol\qty(\Ge^2) .
}{13}
Equation~(\ref{13}) yields $A_{-} = 6$, when $\Ge \to 0$, while $A_{+}$ becomes singular. Hence, the $A_{+}$-branch of the dual flows blows up when $\Ge \to 0$, while the $A_{-}$-branch yields
\eqn{
u_b = 6\, \Pe\, z \, \qty(1 - z)\, \cos\varphi \qc v_b = 6\, \Pe\, z \, \qty(1 - z)\, \sin\varphi \qc w_b = 0, \nonumber\\
T_b = 0 \qc
\grad{p_b} = \Big( - 12\, \Pe\, \cos\varphi ,\ - 12\, \Pe\, \sin\varphi ,\
0 \Big) .
}{14}
Equation~(\ref{14}) represents the isothermal Poiseuille flow in the direction defined by the unit vector $\qty(\cos\varphi, \sin\varphi, 0)$. 

We note that, in every case where $\Ge\ne 0$, the basic temperature gradient is inclined to the horizontal with constant horizontal components in the $x$ and $y$ directions. This feature is easily spotted by reckoning from \equa{9} that $F'''(z) = 12\, \qty(A - 6)$. 

Figures~\ref{fig1} and \ref{fig2} illustrate the velocity profiles as plots of $F(z)$ for various Gebhart numbers and for both branches $A = A_{-}$ and $A = A_{+}$. The $A = A_{-}$ branch shows very small departures from the Poiseuille profile unless the Gebhart number becomes huge, namely, close to its upper bound $\Ge = \sqrt{15}$. This feature is clearly visible in Fig.~\ref{fig1}, while Fig.~\ref{fig2} reveals that the $A = A_{+}$ branch marks a sharp departure from the Poiseuille profile especially at small Gebhart numbers. The characteristics of the $A = A_{+}$ profiles show a bidirectional nature of the flow which arises also, albeit in a minimal way, for the $A = A_{-}$ profiles, but only when $\Ge$ is very close to its upper bound $\sqrt{15}$. Figure~\ref{fig2} also shows that, at small Gebhart numbers, $F(z)$ for the $A = A_{+}$ branch scales proportionally to $\Ge^{-1}$, approximately. This feature is easily recognised on comparing the left-hand frame and the right-hand frame of Fig.~\ref{fig2}, and it is proved on expanding $F(z)$ in a power series of $\Ge$ for $A = A_{+}$,
\eqn{
F(z) = \frac{60\, z\, (1 - z)\, (1 - 2 z)}{\Ge} + 6\, (1 - z)\, z + \ordersymbol\qty(\Ge) .
}{15}
The vertical change of the basic temperature field is illustrated in Figs.~\ref{fig3} and \ref{fig4} by plotting function $G(z)$ with different values of $\Ge$ for the $A = A_{-}$ and $A = A_{+}$ branches. Figure~\ref{fig3} shows that the $A = A_{-}$ branch features a significant dependence of the temperature profiles on $\Ge$, despite the weak influence of the Gebhart number on the velocity profiles as already pointed out while commenting on Fig.~\ref{fig1}. Figure~\ref{fig4} shows that, at small $\Ge$ and for the $A = A_{+}$ branch, function $G(z)$ scales proportionally to $\Ge^{-2}$, approximately. This behaviour is shown through a power series expansion of $G(z)$ with $A = A_{+}$,
\eqn{
G(z) = \frac{720\, z^3\, \qty(6\, z^2 - 15\, z + 10)}{\Ge^2} - \frac{360\, z^2\, (1 - z)^2\, \qty(12\, z^2 - 12\, z + 5)}{\Ge} \nonumber\\
\hspace{3cm}+\, 72\, z^2\, \qty(4\, z^3 - 10\, z^2 + 10\, z - 5) + \ordersymbol\qty(\Ge) .
}{16}
A remarkable feature revealed by Fig.~\ref{fig3} is the unstable thermal stratification $(\partial T_b/\partial z < 0)$ for the $A = A_{-}$ branch occurring at the lower part of the channel. Such an unstable thermal stratification does emerge for all Gebhart numbers also with the $A = A_{+}$ branch, although it can be visualised in Fig.~\ref{fig4} only through the miniatures at small values of $z$. Analytically, this conclusion can be inferred by evaluating $G''(0)$ with $A = A_{+}$,
\eqn{
G''(0) = -\; \frac{24 \left[\qty(\Ge + 5)\, \sqrt{15\, \qty(15 - \Ge^2)} + \Ge\, \qty(15 - \Ge) + 75\right]}{\Ge} .
}{17}
The right hand side of \equa{17} is evidently negative for every $\Ge \le \sqrt{15}$. As a consequence, there always exist regions close to $z=0$ where $G'(z) < 0$.

The existence of a region inside the channel with a negative $z$ component of the basic temperature gradient for either the $A = A_{-}$ branch or the $A = A_{+}$ branch discloses the possibility of a thermal instability of the Rayleigh-B\'enard type (heating from below) for the dual basic flows. This circumstance will be explored later on. It is to be mentioned that the vertical component of the basic temperature gradient is not the only source of a possible thermal instability. In fact, the constant horizontal components of $\grad T_b$ may contribute possibly leading to a thermal instability of the Hadley-type.

It has been pointed out that the $A = A_{+}$ branch can hardly be considered compatible with the Oberbeck-Boussinesq approximate model underlying the existence of the dual solutions \cite{barletta2022mixed}.  The reason is that the approximate model requires the product between the thermal expansion coefficient $\beta$ and the maximum temperature difference across the flow domain to be much smaller than unity. As pointed out by \citet{barletta2022mixed}, this condition is precluded when considering the $A = A_{+}$ branch of the dual solutions. On account of these findings, the forthcoming stability analysis will be focussed just on the $A = A_{-}$ branch.

\section{Dynamics of small-amplitude perturbations}
Let us perturb the basic dual flows,
\eqn{
\pmqty{\vb{u}\\p\\T} = \pmqty{\vb{u}_b\\p_b\\T_b} + \varepsilon\! \pmqty{\vb{U}\\P\\\Theta} ,
}{18}
where $\varepsilon$ is the perturbation parameter and $\qty(\vb{U},P,\Theta)$ are the perturbations of velocity, pressure and temperature, respectively. The Cartesian components of $\vb{U}$ are denoted as $\qty( U,V,W )$. Let us substitute \equa{18} into \equasa{5}{6} by employing \equasa{2}{8}, namely
\gat{
\div{\vb{U}} = 0, \label{19a}\\
\frac{1}{\Pr} \left[ \pdv{\vb{U}}{t} + (\vb{u}_b \cdot \grad{)\,\vb{U}} + (\vb{U} \cdot \grad{)\,\vb{u}_b} \right] = - \grad{P} + \Theta\, \vu{e}_z + \nabla^2 \vb{U},\label{19b}\\
\pdv{\Theta}{t} + (\vb{u}_b \cdot \grad{)\,\Theta} + (\vb{U} \cdot \grad{)\,T_b} = \nabla^2 \Theta + 2\, \Ge\, \qty[ \qty(\pdv{U}{z} + \pdv{W}{x})\, u'_b + \qty(\pdv{V}{z} + \pdv{W}{y})\, v'_b ] , \label{19c}\\
\vb{U} = 0 \qc \pdv{\Theta}{z} = 0 \qfor z = 0, 1, \label{19d}
}{19}
where terms $\ordersymbol(\varepsilon^2)$ have been neglected in order to account for small-amplitude perturbations.

Since the basic flow direction is inclined an angle $\varphi$ to the $x$ direction, the $x$ axis defines an arbitrary horizontal direction, so that the normal modes of perturbation can be expressed as plane waves propagating along the $x$-direction. The effect of arbitrary oblique modes can be tested by allowing a changing angle within the range $0 \le \varphi \le \pi/2$. The values $\varphi=0$ and $\varphi=\pi/2$ yield the transverse modes and the longitudinal modes, respectively. Thus, we write
\eqn{
\pmqty{\vb{U}\\P\\\Theta} = \pmqty{\hat{\vb{U}}(z) \\ \hat{P}(z) \\ \hat{\Theta}(z)} e^{i \,k\, x} \, e^{\lambda\, t},
}{20}
where $k$ is the real-valued wavenumber, while $\lambda$ is a complex-valued parameter. The real and imaginary parts of $\lambda$ are denoted as $\lambda = \eta - i\, \omega$, with $\eta$ expressing the growth rate and $\omega$ yielding the angular frequency. Linear instability is identified with the condition $\eta > 0$, while $\eta = 0$ expresses the threshold case of neutral stability. We substitute \equa{20} into \equas{19}, so that we can write
\gat{
i\, k\, \hat{U} + \hat{W}' = 0, \label{21a}\\
\frac{1}{\Pr} \left( \lambda\, \hat{U} + i\, k\, u_b\, \hat{U} + \hat{W}\, u'_b \right) = - i\, k\, \hat{P} + \hat{U}'' - k^2\, \hat{U} ,\label{21b}\\
\frac{1}{\Pr} \left( \lambda\, \hat{V} + i\, k\, u_b\, \hat{V} + \hat{W}\, v'_b \right) = \hat{V}'' - k^2\, \hat{V},\label{21c}\\
\displaybreak
\frac{1}{\Pr} \left( \lambda\, \hat{W} + i\, k\, u_b\, \hat{W} \right) = - \hat{P}' + \hat{\Theta} + \hat{W}'' - k^2\, \hat{W} ,\label{21d}\\
\lambda\, \hat{\Theta} + i\, k\, u_b\, \hat{\Theta} + \hat{U}\, \pdv{T_b}{x} + \hat{V}\, \pdv{T_b}{y} + \hat{W}\, \pdv{T_b}{z} \nonumber\\
\hspace{5cm}= \hat{\Theta}'' - k^2\, \hat{\Theta} + 2\, \Ge\, \qty[ \qty(\hat{U}' + i\,k\, \hat{W})\, u'_b + \hat{V}'\, v'_b ] , \label{21e}\\
\hat{U} = 0 \qc \hat{V} = 0 \qc \hat{W} = 0 \qc \hat{\Theta}' = 0 \qfor z = 0, 1, \label{21f}
}{21}
where $(\hat{U}, \hat{V}, \hat{W})$ are the Cartesian components of $\hat{\vb{U}}$.

The physically significant situation where the viscous dissipation effect is expected to cause the instability is when the fluid has a large kinematic viscosity combined with a small thermal diffusivity. Roughly speaking, such fluids are markedly susceptible to frictional heating while the diffusion of such an internally generated heat is inefficient. This situation reasonably occurs when the Prandtl number is extremely large.

\subsection{Creeping Flow}
A regime of creeping flow occurs when the inertial terms in the momentum balance are negligible. This happens when the limit $\Pr \to \infty$ is taken for \equa{19b}. As mentioned above, such a limit identifies a condition where an extremely viscous fluid is employed having a small thermal diffusivity. In this limit, \equas{21} simplify to
\gat{
i\, k\, \hat{U} + \hat{W}' = 0, \label{22a}\\
- i\, k\, \hat{P} + \hat{U}'' - k^2\, \hat{U} = 0,\label{22b}\\
\hat{V}'' - k^2\, \hat{V} = 0,\label{22c}\\
- \hat{P}' + \hat{\Theta} + \hat{W}'' - k^2\, \hat{W} = 0 ,\label{22d}\\
\lambda\, \hat{\Theta} + i\, k\, u_b\, \hat{\Theta} + \hat{U}\, \pdv{T_b}{x} + \hat{V}\, \pdv{T_b}{y} + \hat{W}\, \pdv{T_b}{z} \nonumber\\
\hspace{5cm}= \hat{\Theta}'' - k^2\, \hat{\Theta} +\, 2\, \Ge\, \qty[ \qty(\hat{U}' + i\,k\, \hat{W})\, u'_b + \hat{V}'\, v'_b ] , \label{22e}\\
\hat{U} = 0 \qc \hat{V} = 0 \qc \hat{W} = 0 \qc \hat{\Theta}' = 0 \qfor z = 0, 1. \label{22f}
}{22}
From \equasa{22c}{22f}, one can immediately conclude that $\hat{V}=0$ in the whole range $0 \le z \le 1$. Furthermore, by employing \equasa{22a}{22b}, one can express $\hat{U}$ and $\hat{P}$ in terms of $\hat{W}$ and its derivatives. Thus, one is led to a reformulation of \equas{22} involving only the unknowns $\hat{W}$ and $\hat{\Theta}$,
\gat{
\hat{W}'''' - 2\,k^2\, \hat{W}'' + k^4\, \hat{W} - k^2\, \hat{\Theta} = 0 ,\label{23a}\\
\hat{\Theta}'' - \qty[k^2 + \lambda + i\, k\, \Pe\, F(z) \cos\varphi ]\, \hat{\Theta} + \frac{2\, i\, \Ge\, \Pe\, F'(z) \cos\varphi}{k}\, \qty(\hat{W}'' + k^2\, \hat{W}) \nonumber\\
\hspace{3cm}- \frac{i\, \Pe\, F'''(z) \cos\varphi}{k}\, \hat{W}' - \Pe^2\, G'(z)\, \hat{W} = 0  , \label{23b}\\
\hat{W} = 0 \qc \hat{W}' = 0 \qc \hat{\Theta}' = 0 \qfor z = 0, 1. \label{23c}
}{23}
Creeping flow is usually associated with a regime of extremely small Reynolds number, $\Rey = \Pe/\Pr$. From the mathematical viewpoint, creeping flow means a situation where the limit $\Rey \to 0$ is combined with the limit $\Pr \to \infty$. These limits can be taken on keeping $\Pe = \Rey\, \Pr \sim {\rm O}(1)$. In the following, we will consider the dynamics of perturbations as modelled by a creeping flow achieved with a finite P\'eclet number.

The linear analysis of the instability for the basic flows defined by \equass{8}{10} is carried out by solving numerically the eigenvalue problem (\ref{23}). The solution is sought by fixing the input parameters $(\varphi, \Ge)$ and by determining the neutral stability threshold value of $\Pe$ versus the wavenumber $k$. The angular frequency, $\omega$, of the neutrally stable modes is also determined. 

The neutral stability threshold implies a zero growth rate, $\eta$. Additionally, for the sake of convenience, the basic average velocity is taken into account by redefining the angular frequency in the comoving reference frame, namely
\eqn{
\hat\omega = \omega - k\, \Pe\, \cos\varphi .
}{24}
Thus, \equas{23} can be rewritten as
\gat{
\hat{W}'''' - 2\,k^2\, \hat{W}'' + k^4\, \hat{W} - k^2\, \hat{\Theta} = 0 ,\label{25a}\\
\hat{\Theta}'' - \qty[k^2 - i\, \hat{\omega} + i\, k\, \Pe\, \hat{F}(z) \cos\varphi ]\, \hat{\Theta} + \frac{2\, i\, \Ge\, \Pe\, \hat{F}'(z) \cos\varphi}{k}\, \qty(\hat{W}'' + k^2\, \hat{W}) \nonumber\\
\hspace{3cm}- \frac{i\, \Pe\, \hat{F}'''(z) \cos\varphi}{k}\, \hat{W}' - \Pe^2\, G'(z)\, \hat{W} = 0  , \label{25b}\\
\hat{W} = 0 \qc \hat{W}' = 0 \qc \hat{\Theta}' = 0 \qfor z = 0, 1 ,\label{25c}
}{25}
where $\hat{F}(z)=F(z)-1$ is a function with a zero average value over the interval $0 \le z \le 1$, as a consequence of \equa{12}. By assigning $\varphi$ as an input datum for the solution of \equas{25}, one actually defines the type of oblique modes perturbing the flow, with the transverse modes $(\varphi=0)$ and the longitudinal modes $(\varphi=\pi/2)$ as limiting cases.

\section{Discussion of the Results}
We pointed out in Section~\ref{badufl} that the linear stability analysis is relative to the $A=A_-$ branch of the dual flows. The numerical solution of the stability eigenvalue problem (\ref{25}) is sought by employing the shooting method. We do not go into the details of this numerical technique as its use for the solution of flow stability eigenvalue problems has been widely discussed elsewhere \cite{straughan2013energy, barletta2019routes}. 

\begin{figure}[ht!]
\centering
\includegraphics[width=0.999\textwidth]{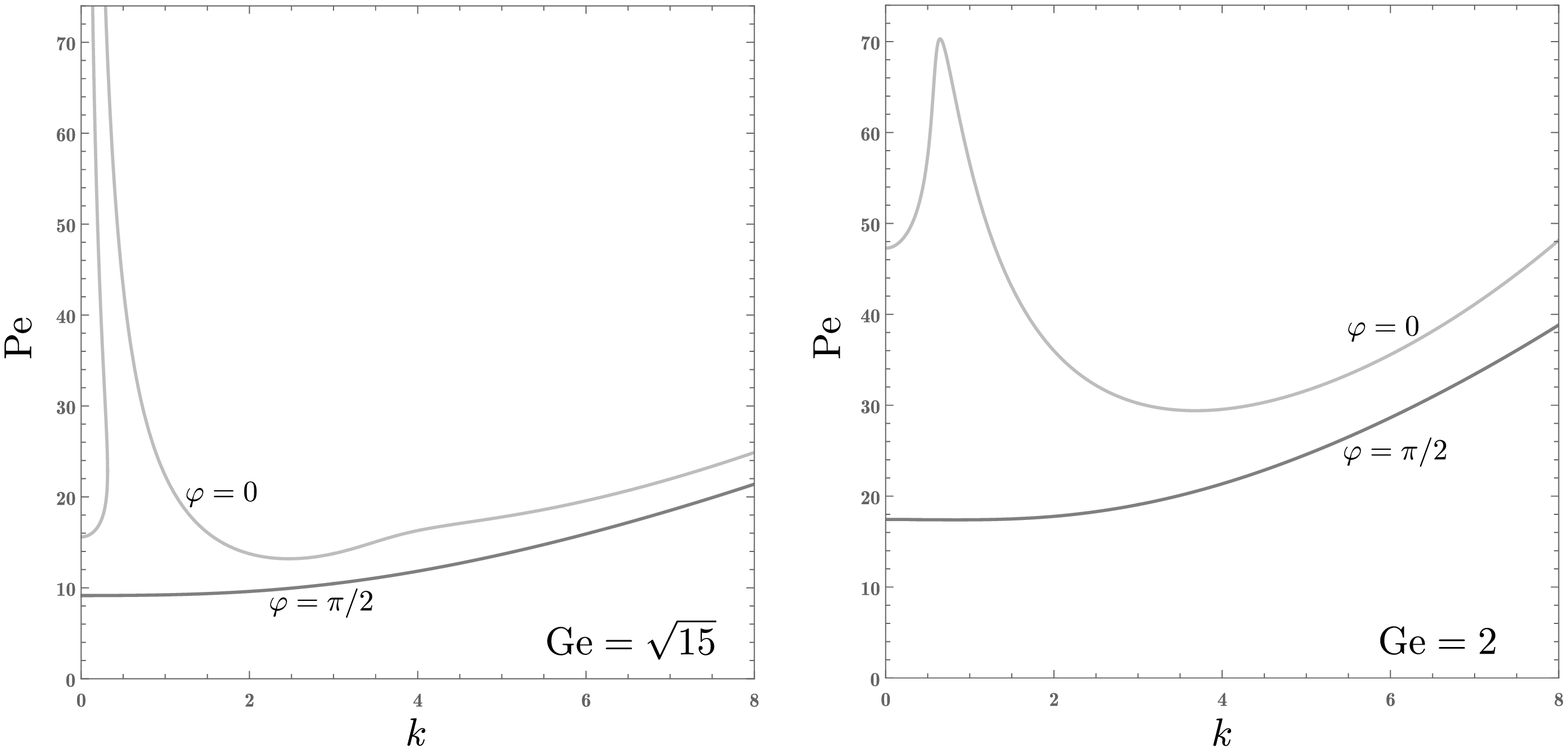}
\caption{\label{fig5}Neutral stability curves in the $(k, \Pe)$ plane for transverse (light grey line, $\varphi=0$) and longitudinal (dark grey line, $\varphi=\pi/2$) modes}
\end{figure}

\begin{figure}[h!]
\centering
\includegraphics[width=\textwidth]{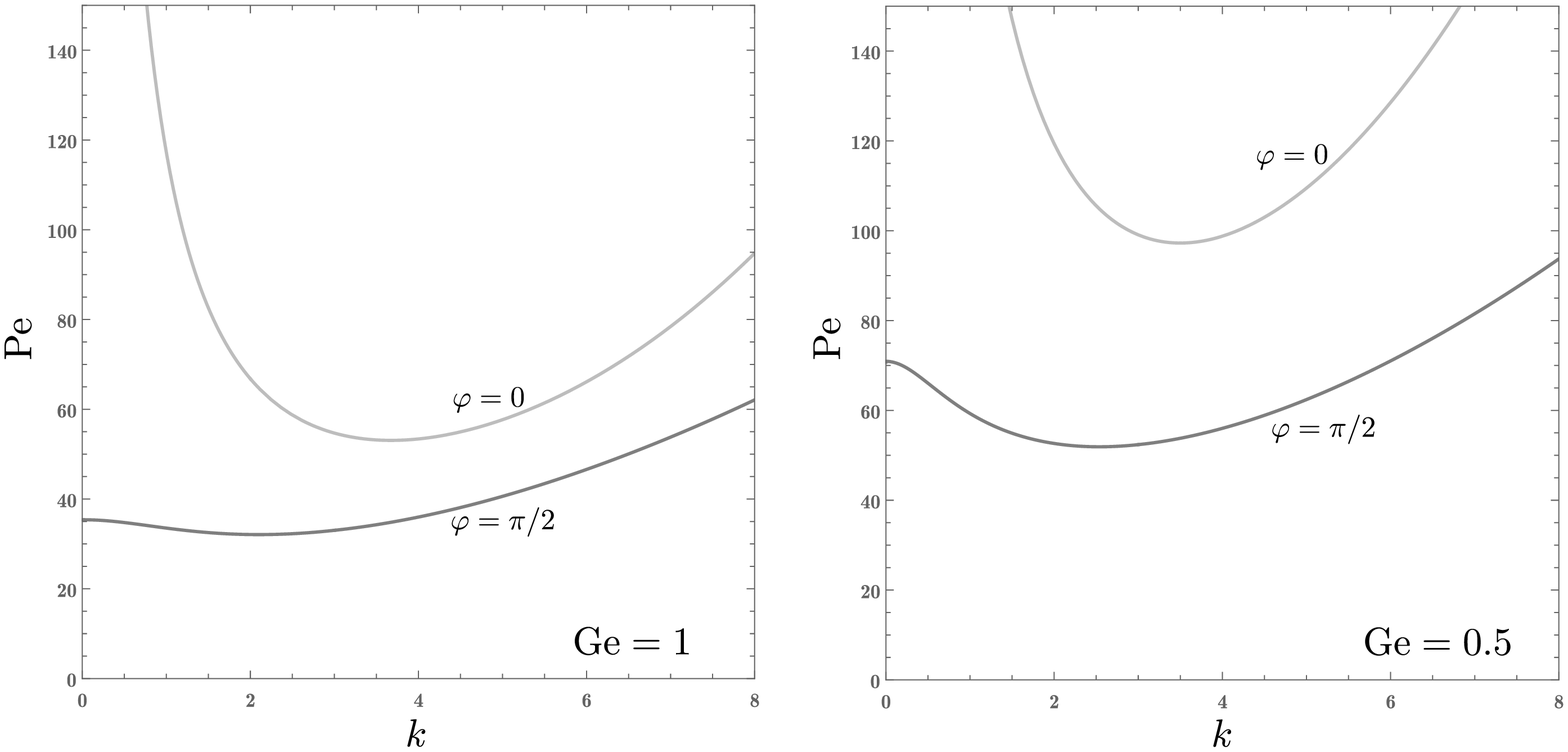}
\caption{\label{fig6}Neutral stability curves in the $(k, \Pe)$ plane for transverse (light grey line, $\varphi=0$) and longitudinal (dark grey line, $\varphi=\pi/2$) modes}
\end{figure}

The software tool actually employed to implement the shooting method is {\sl Mathematica} (\copyright{~}Wolfram Research, Inc.) with its functions {\tt NDSolve} and {\tt FindRoot}. The former function serves to solve the initial value problem based on \equas{25}, starting at $z=0$, while the latter function solves the target conditions (\ref{25c}) at $z=1$, thus yielding the numerical values for the neutral stability output data $(\hat\omega, \Pe)$. The neutral stability condition is represented by the curve in the parametric $(k, \Pe)$ plane, with the minimum $\Pe$ point along the curve yielding the critical condition for the onset of instability. Such a critical condition is given by $k=k_c$ and $\Pe = \Pe_c$, where the subscript $c$ stands for critical value. Tracing the dependence of $\Pe_c$ on the inclination angle $\varphi$ of the perturbation mode in different cases allows one to establish which modes are the most effective at onset of instability. Then, the stability analysis can be focussed just on those modes.

\subsection{The Most Unstable Perturbations}
We start the stability analysis from high values of $\Ge$, namely $\Ge = \sqrt{15}$ and $\Ge=2$. The neutral stability curves in the $(k, \Pe)$ plane are drawn in Fig.~\ref{fig5} for the transverse and the longitudinal modes. We point out that $\Ge = \sqrt{15}$ is the highest possible value of the Gebhart number. In fact, such a value is the highest possible according to the mathematical definition of the dual solutions, but we stress that this value is extremely large for physical systems. Figure~\ref{fig5} reveals that, both for $\Ge = \sqrt{15}$ and $\Ge=2$, the longitudinal modes are the more unstable as they yield the transition to instability with lower values of $\Pe$ for every $k$. The neutral stability curves for longitudinal modes reveal also that a finite value of $\Pe$ is achieved when $k \to 0$. This feature has been observed also in the study of the Rayleigh-B\'enard instability when Neumann boundary conditions for the temperature are utilised, instead of the usual Dirichlet temperature conditions \cite{park1991hydrodynamic}. We point out that the critical value of $\Pe$ for longitudinal modes  is obtained for $k \to 0$ in the case $\Ge = \sqrt{15}$, but for a nonzero $k$ with $\Ge = 2$ though hardly evident in Fig.~\ref{fig5}. More precisely, one finds
\eqn{
k_c = 0.842235 \qc \Pe_c = 17.4116 \qc \qfor \Ge = 2 .
}{26}
We also note that the neutral stability curve for transverse modes is disconnected  in two parts when $\Ge = \sqrt{15}$. These disconnected parts join in a single curve when $\Ge =2$, thus forming a local maximum of $\Pe$ versus $k$. 

\begin{figure}[t]
\centering
\includegraphics[width=\textwidth]{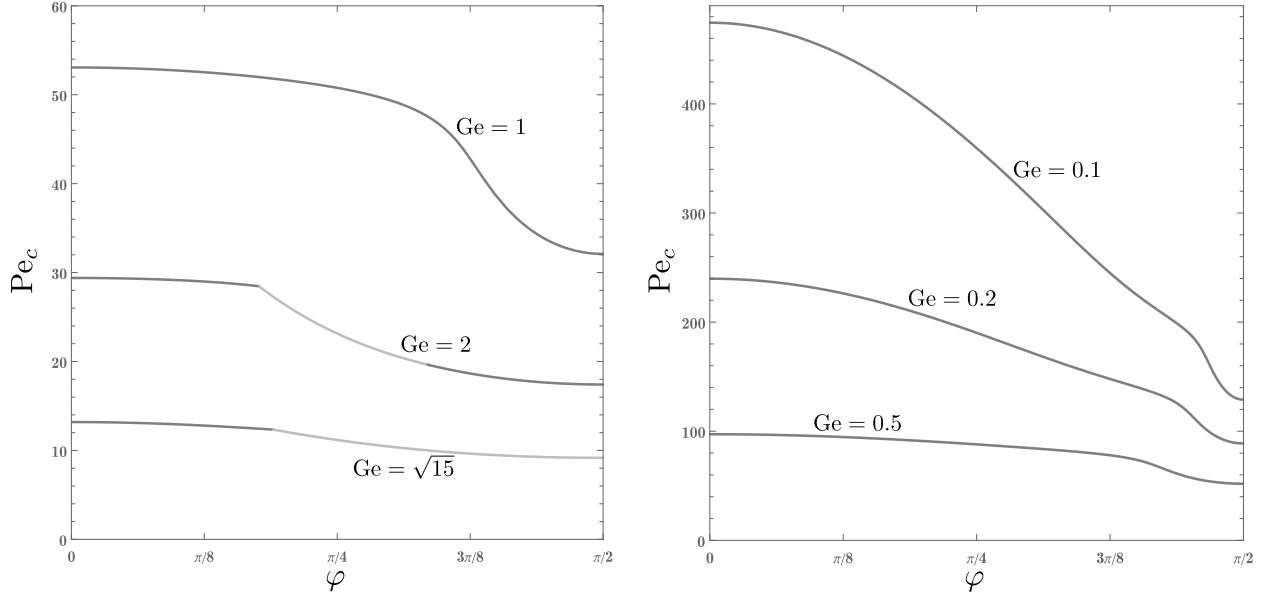}
\caption{\label{fig7}Critical values of $\Pe$ versus $\varphi$ for oblique modes. Light grey lines denote $k_c=0$ critical conditions, while dark grey lines denote $k_c\ne 0$ critical conditions}
\end{figure}

The conclusion that longitudinal modes are the most unstable is confirmed by Fig.~\ref{fig6} which is relative to $\Ge = 1$ and $\Ge = 0.5$. For both values of $\Ge$, the critical condition $\Pe = \Pe_c$ occurs for longitudinal modes with a nonzero $k_c$. There is a markedly different shape of the neutral stability curves for transverse modes with $\Ge = 1$ and $\Ge = 0.5$ compared to that for $\Ge = 2$ (Fig.~\ref{fig5}). Such a diversion is due to the very steep increase of the neutral stability value of $\Pe$ in the limit $k \to 0$ when $\Ge$ decreases below 2. 

A more systematic study of the effects of the inclination angle, $\varphi$, on the instability threshold is carried out by testing the dependence of the critical P\'eclet number on this angle. Together with the huge values of $\Ge$ considered in Figs.~\ref{fig5} and \ref{fig6}, smaller and gradually more realistic values of $\Ge$ are considered in Fig.~\ref{fig7}. This figure shows the change of $\Pe_c$ with $\varphi$ for oblique modes. The change is monotonic with the case $\varphi=\pi/2$ (longitudinal modes) displaying the minimum value of $\Pe_c$, in all case examined. This means that the statement that longitudinal modes are those selected at onset of instability can be assumed of a general validity. Figure~\ref{fig7} reveals that the sensitivity of $\Pe_c$ to $\varphi$ becomes stronger as $\Ge$ decreases. Another important fact is that the critical value of $\Pe$ is achieved with either $k_c = 0$ or $k_c \ne 0$ for different ranges of $\varphi$. Indeed, such a complicated trend is a peculiarity for those cases where $\Ge$ is very large, $\sqrt{15}$ or $2$ among the values examined in Fig.~\ref{fig7}. For smaller Gebhart numbers, the critical conditions are always associated with $k_c \ne 0$ for every $\varphi$.

\begin{figure}[t]
\centering
\includegraphics[width=0.5\textwidth]{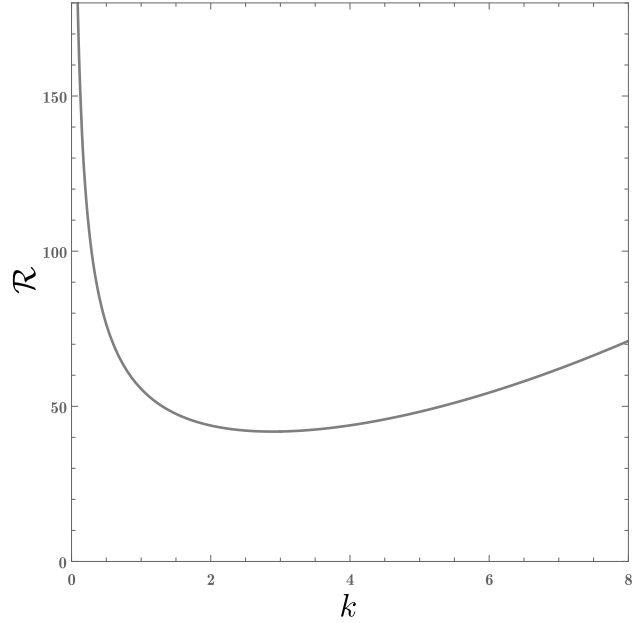}
\caption{\label{fig8}Longitudinal modes: neutral stability curve in the $(k, \RR)$ plane for the asymptotic condition $\Ge \to 0$}
\end{figure}

\subsection{The Longitudinal Modes}
Since we established that the longitudinal modes identify the onset of the instability, we will now restrict our attention to these modes. Then, the stability eigenvalue problem (\ref{25}) is greatly simplified,
\gat{
\hat{W}'''' - 2\,k^2\, \hat{W}'' + k^4\, \hat{W} - k^2\, \hat{\Theta} = 0 ,\label{27a}\\
\hat{\Theta}'' - \qty( k^2 - i\, \hat{\omega} )\, \hat{\Theta} - \Pe^2\, G'(z)\, \hat{W} = 0  , \label{27b}\\
\hat{W} = 0 \qc \hat{W}' = 0 \qc \hat{\Theta}' = 0 \qfor z = 0, 1 .     \label{27c}
}{27}
The most important feature of \equas{27} is that neutral stability with longitudinal modes occurs with $\hat\omega = \omega = 0$, where the equality $\hat\omega = \omega$ is a consequence of the definition (\ref{24}). This feature, which can be designated as a principle of exchange of stabilities, cannot be proved rigorously though it can be inferred quite clearly from the numerical data. 

\subsubsection{The Limit of Small Gebhart Numbers}\label{smaGenum}
A physically significant asymptotic condition is associated with the double limit $\Ge \to 0$ and $\Pe \to \infty$ on keeping $\Ge\, \Pe^2 \sim \order{1}$. Beyond its mathematical formulation, this asymptotic behaviour means that with small values of $\Ge$ the neutral stability threshold for $\Pe$ scales proportionally to $\Ge^{-1/2}$.

By expanding $G'(z)$ in a power series of $\Ge$, on account of \equasa{9}{10} and of the choice $A = A_-$, we obtain
\eqn{
G'(z) = - 36 \, \Ge\, \qty( 2 z^3 - 3 z^2 + z ) + \order{\Ge^2}.
}{28}
We also introduce the parameter
\eqn{
\RR = \Pe \, \sqrt{\Ge} .
}{29}
Thus, with $\Ge \to 0$ and $\RR \sim \order{1}$, \equasa{27}{28} yield
\gat{
\hat{W}'''' - 2\,k^2\, \hat{W}'' + k^4\, \hat{W} - k^2\, \hat{\Theta} = 0 ,\label{30a}\\
\hat{\Theta}'' - \qty( k^2 - i\, \hat{\omega} )\, \hat{\Theta} + 36\, \RR^2\,  \qty( 2 z^3 - 3 z^2 + z )\, \hat{W} = 0  , \label{30b}\\
\hat{W} = 0 \qc \hat{W}' = 0 \qc \hat{\Theta}' = 0 \qfor z = 0, 1 .     \label{30c}
}{30}
We note that, in \equas{30}, $\Ge$ and $\Pe$ do not appear separately but only through the parameter $\RR$. Hence, the neutral stability in the limit $\Ge \to 0$ can be formulated as a threshold condition for the parameter $\RR$. 
We also point out that $\RR^2$ can be interpreted as a viscous dissipation based Rayleigh number
\eqn{
\RR^2 = \frac{g\, \beta\, \Delta T_{vd}\, H^3}{\nu \alpha} \qc \text{where} \quad \Delta T_{vd} = \frac{\mu\, U_0^2}{\chi} ,
}{31}
and \equasa{7}{11} have been used. Here, $\mu$ is the dynamic viscosity of the fluid and $\chi$ its thermal conductivity, while $\Delta T_{vd}$ is a dimensional temperature difference characteristic of the viscous dissipation effect.

The numerical solution of \equas{30} allows one to obtain the neutral stability curve in the $(k, \RR)$ plane for the limiting case of small-$\Ge$, as illustrated in Fig.~\ref{fig8}. An important feature is that, unlike the neutral stability curves for finite nonzero values of $\Ge$ shown in Figs.~\ref{fig5} and \ref{fig6}, the ordinate axis $k=0$ is a vertical asymptote for the neutral stability curve shown in Fig.~\ref{fig8}. In fact, the critical data for the small-$\Ge$ asymptotic solution are
\eqn{
k_c = 2.88872 \qc \RR_c = 41.8534
}{32}
Equation~(\ref{32}) is the basis for determining the trend of $\Pe_c$ versus $\Ge$ at small values of $\Ge$. In particular, \equasa{29}{32} imply that $\Pe_c \sim {\cal O}(\Ge^{-1/2})$. This result is interesting as it differs significantly from the behaviour observed in the case of the adiabatic Darcy's flow in a porous medium \cite{barletta2009stability}. In that case, it has been proved by \citet{barletta2009stability} that the small-$\Ge$ trend of $\Pe_c$ is such that $\Pe_c \sim {\cal O}(\Ge^{-1})$. In other words, for Darcy's flow in a porous medium, $\Pe_c$ diverges to infinity for $\Ge \to 0$ significantly faster than in the case of Newtonian flow. The reason of the difference between these cases relies in the derivative $\partial T_b / \partial z$, which drives the instability for longitudinal modes. As $\Ge \to 0$, one can infer from \equas{8}, (\ref{9}) and (\ref{10}) with $A = A_{-}$ that $\partial T_b / \partial z \sim {\cal O}(\Ge\, \Pe^2)$. The corresponding result in the case of Darcy's flow is $\partial T_b / \partial z \sim {\cal O}(\Ge^2\, \Pe^2)$ as demonstrated in \citet{barletta2009stability}.

\subsubsection{The Neutral Stability Condition for $k \to 0$}\label{smak}
It has been pointed out that, for every $\Ge \ne 0$, the neutral stability condition for longitudinal modes suggests a finite value of $\Pe$ for infinite wavelenght modes, {\em i.e.}, for the limit $k \to 0$. This conjectured trend can be verified analytically by employing a power series solution of the eigenvalue problem (\ref{27}).
We set $\hat\omega = 0$ and adopt the expansions in even powers of $k$ given by
\eqn{
\hat{W}(z) = \hat{W}_0(z) + \hat{W}_2(z) \, k^2 + \hat{W}_4(z) \, k^4 + \order{k^6},
\nonumber\\
\hat{\Theta}(z) = \hat{\Theta}_0(z) + \hat{\Theta}_2(z) \, k^2 + \hat{\Theta}_4(z) \, k^4 + \order{k^6},
\nonumber\\
\Pe = \Pe_0 + \Pe_2 \, k^2 + \Pe_4 \, k^4 + \order{k^6}.
}{33}
By substituting \equa{33} into \equas{27} we obtain, to zero order in $k$,
\gat{
\hat{W}_0'''' = 0 ,\label{34a}\\
\hat{\Theta}_0'' - \Pe_0^2\, G'(z)\, \hat{W}_0 = 0  , \label{34b}\\
\hat{W}_0(0) = 0 \qc \hat{W}_0'(0) = 0 \qc \hat{\Theta}_0'(0) = 0 ,    \nonumber\\
\hat{W}_0(1) = 0 \qc \hat{W}_0'(1) = 0 \qc \hat{\Theta}_0'(1) = 0 .     \label{34c}
}{34}
The solution of \equas{34} is
\eqn{
\hat{W}_0(z) = 0 \qc \hat{\Theta}_0(z) = 1,
}{35}
while $\Pe_0$ remains yet undetermined. 

\begin{figure}[t]
\centering
\includegraphics[width=0.5\textwidth]{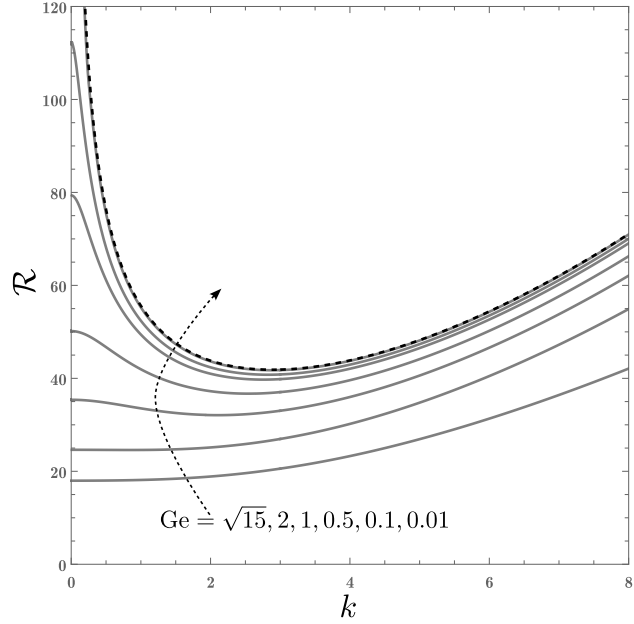}
\caption{\label{fig9}Longitudinal modes: neutral stability curves in the $(k, \RR)$ plane for different values of $\Ge$. The dashed neutral stability curve indicates the asymptotic condition $\Ge \to 0$}
\end{figure}

\begin{figure}[h!]
\centering
\includegraphics[width=0.5\textwidth]{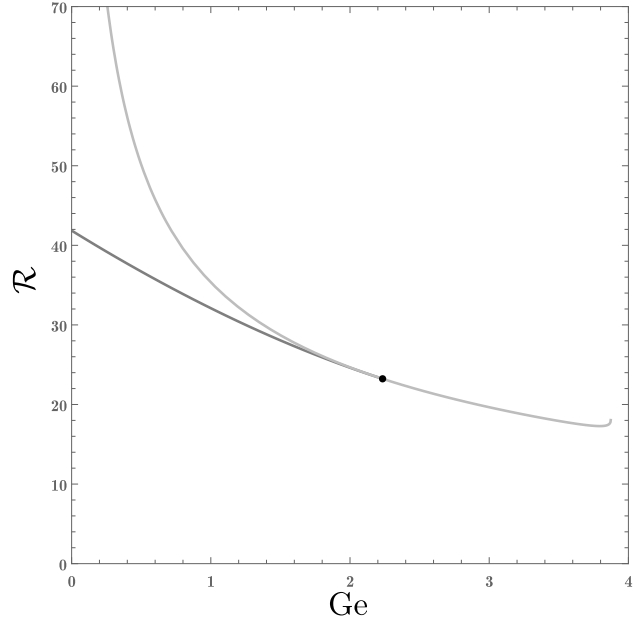}
\caption{\label{fig10}Longitudinal modes: $\RR_c$ versus $\Ge$ (dark grey line) compared with $\RR_0$ versus $\Ge$ (light grey line). The black dot denotes the transition from $k_c \ne 0$ to $k_c = 0$ occurring when $\Ge$ is given by \equa{36rees2}}
\end{figure}

Indeed, $\hat{\Theta}_0(z)$ could be equal to any constant value as the original problem (\ref{27}) is homogeneous. 
Choosing the constant value $1$ in \equa{35} means getting rid of the scale invariance for the eigenfunctions by implicitly fixing the extra condition $\hat{\Theta}(0) = 1$. The immediate consequence is that $\hat{\Theta}_n(0) = 0$ for every positive integer $n$. Therefore, the system obtained from \equas{27} to second order in $k$ is given by
\gat{
\hat{W}_2'''' - 1 = 0 ,\label{36a}\\
\hat{\Theta}_2'' - 1 - \Pe_0^2\, G'(z)\, \hat{W}_2 = 0  , \label{36b}\\
\hat{W}_2(0) = 0 \qc \hat{W}_2'(0) = 0 \qc \hat{\Theta}_2(0) = 0 \qc \hat{\Theta}_2'(0) = 0 ,    \nonumber\\
\hat{W}_2(1) = 0 \qc \hat{W}_2'(1) = 0  .     \label{34c}
}{36}
There should also be an extra condition $\hat{\Theta}_2'(1) = 0$, but it is not needed to determine the solution of \equas{36}. We do not report such a solution for the sake of brevity. We just say that $\hat{W}_2(z)$ is a fourth-degree polynomial in $z$, while $\hat{\Theta}_2(z)$ is an eleventh-degree polynomial in $z$. 

Forcing the extra condition $\hat{\Theta}_2'(1)=0$ results in a relation between $\Pe_0$ and $\Ge$, namely
\eqn{
\Pe_0 = 2 \sqrt{\frac{105\, \Ge^2}{\qty(20 - \Ge^2) \sqrt{15 \qty(15-\Ge^2)} + 25\, \Ge^2 - 300}} .
}{37}
A characteristic feature of \equa{37} is that $Pe_0 \sim \order{\Ge^{-2}}$ in the limit $\Ge \to 0$. This feature justifies our statement in Section~\ref{smaGenum} that the ordinate axis $k=0$ is a vertical asymptote for the neutral stability curve when $\Ge \to 0$. Equation~\ref{37} provides also a rigorous evaluation of the critical P\'eclet number in the maximum-$\Ge$ case, {\em i.e.}, for $\Ge = \sqrt{15}$. In fact, $k_c = 0$ in this case and
\eqn{
\Pe_c = \Pe_0 = 2 \sqrt{21} \approx 9.16515 .
}{38}
By comparing this result with the numerical value of $\Pe$ along the neutral stability curve (see Fig.~\ref{fig5}) with $k=0.01$, we estimate a relative discrepancy less than $0.01\, \%$. Such a discrepancy is extremely satisfactory, given that the numerical shooting-method solver employed for the eigenvalue problem (\ref{27}) yields the smallest accuracy when $k$ is close to $0$.

By considering the system obtained from \equas{27} to fourth order in $k$, one obtains
\gat{
\hat{W}_4'''' - 2\, \hat{W}_2'' - \hat{\Theta}_2 = 0 ,\label{36reesa}\\
\hat{\Theta}_4'' - \hat\Theta_2 - \qty(\Pe_0^2\, \hat{W}_4 + 2\, \Pe_0\, \Pe_2\, \hat{W}_2)\, G'(z) = 0  , \label{36reesb}\\
\hat{W}_4(0) = 0 \qc \hat{W}_4'(0) = 0 \qc \hat{\Theta}_4(0) = 0 \qc \hat{\Theta}_4'(0) = 0 ,    \nonumber\\
\hat{W}_4(1) = 0 \qc \hat{W}_4'(1) = 0  .     \label{34reesc}
}{36rees}
The extra condition $\hat{\Theta}_4'(1)=0$ yields a relation between $\Pe_2$ and $\Ge$. We do not report the analytical expression of $\Pe_2$, but we just note that it can be used to detect the threshold for $k_c$ to change from zero to nonzero as $\Ge$ increases. We already pointed out that the neutral stability curve for longitudinal modes with $\Ge=2$ has $k_c \ne 0$, while that for $\Ge = \sqrt{15}$ has $k_c = 0$. The threshold value of $\Ge$ where $k_c$ changes from a nonzero to a zero value can be detected as that value corresponding to a change in the concavity of the neutral stability curve at $k=0$. Such a change in concavity corresponds to a change of sign for $\Pe_2$. In fact, the condition $\Pe_2 = 0$ yields
\eqn{
\Ge = 2.23397 .
}{36rees2}

\subsubsection{The Neutral Stability Curves and the Critical Values}
The neutral stability curves for longitudinal modes represented in the $(k, \RR)$ plane are reported in Fig.~\ref{fig9} for decreasing values of $\Ge$ ranging from its maximum, $\sqrt{15}$, to $0.01$. The neutral stability curve for the asymptotic case $\Ge \to 0$ (dashed line) is reported for comparison. The first impression is that the neutral stability curve for $\Ge=0.01$ matches almost perfectly that for $\Ge \to 0$. However, by employing \equa{37}, one should keep in mind that the neutral stability curve for $\Ge = 0.01$ yields a finite, though very large, value of $\RR$ in the limit $k \to 0$, {\em i.e.} $\RR=354.967$, while the neutral stability curve for the asymptotic case $\Ge \to 0$ has a vertical asymptote at $k=0$.

\begin{table}[t]
\centering
\begin{tabular}{ c | rrrrr }
$\Ge$ & $k_c$ & $\Pe_c$ & $\RR_c$ & $\Pe_0$ & $\RR_0$\\
  \hline			
  0      & 2.88872 & $\infty$ & 41.8534 & $\infty$ & $\infty$ \\
  0.01  & 2.88237 & 417.4493 & 41.7449 & 3549.6464 & 354.9646\\
  0.1    & 2.82398 & 128.9560 & 40.7795 & 354.9500 & 112.2450\\
  0.2    & 2.75634 & 88.8359 & 39.7286 & 177.4528 & 79.3593\\
  0.3    & 2.68571 & 70.6563 & 38.7001 & 118.2772 & 64.7831\\
  0.4    & 2.61199 & 59.5981 & 37.6932 & 88.6819 & 56.0874\\
  0.5    & 2.53508 & 51.9123 & 36.7076 & 70.9188 & 50.1472\\
  0.8    & 2.28435 & 37.8763 & 33.8776 & 44.2515 & 39.5797\\
  1       & 2.09952 & 32.0986 & 32.0986 & 35.3470 & 35.3470\\
  1.2    & 1.89924 & 27.7601 & 30.4097 & 29.4002 & 32.2063\\
  1.5    & 1.56474 & 22.9062 & 28.0543 & 23.4372 & 28.7045\\
  1.8    & 1.17077 & 19.3205 & 25.9212 & 19.4448 & 26.0879\\
  2       & 0.84224 & 17.4116 & 24.6237 & 17.4397 & 24.6634\\
  2.2    & 0.31397 & 15.7919 & 23.4232 & 15.7924 & 23.4239\\
  2.23397 & 0 & 15.5412 & 23.2286 & 15.5412 & 23.2286\\
  $\sqrt{15}$ & 0 & 9.1652 & 18.0369 & 9.1652 & 18.0369\\
  \hline  
\end{tabular}
\caption{\label{tab1} Values of $k_c$, $\Pe_c$, $\RR_c$, $\Pe_0$, $\RR_0$ versus $\Ge$}
\end{table}

Figure~\ref{fig10} displays a comparison between the critical value of $\RR$ and the $k \to 0$ limiting value of $\RR$ versus $\Ge$. In fact, for a given $\Ge$, we define
\eqn{
\RR_c = \Pe_c \sqrt{\Ge} \qc \RR_0 = \Pe_0 \sqrt{\Ge}. 
}{39}
A dark grey line is employed for $\RR_c$ and a light grey line for $\RR_0$. The latter line almost overlaps the former when $\Ge$ is greater than $2$, while the exact overlapping occurs when $\Ge$ exceeds the threshold value given by \equa{36rees2} and identified in Fig.~\ref{fig10} by a black dot. The reason is that the difference between $\RR_c$ and $\RR_0$ becomes very small when $\Ge$ is so large. When $\Ge \to 0$, the value of $\RR_c$ agrees with that given by \equa{32}. When $\Ge = \sqrt{15} \approx 3.87298$, both the lines come to an end as this value of $\Ge$ is the maximum possible for the existence of the basic state. Although hardly significant for a physical system, the range of $\Ge$ very close to its maximum shows a diversion from the general decreasing trend of both $\RR_c$ and $\RR_0$ versus $\Ge$ with a minimum which can be accurately evaluated for $\RR_0$ by using \equa{37},
\eqn{
\RR_{0, \min} =  3  \sqrt{21\sqrt{\frac{5}{2}}} \approx 17.2869   \qc           
\Ge_{\min} =  6 \sqrt{\frac{2}{5}} \approx 3.79473   .
}{40}
Table~\ref{tab1} reports the critical data for the parameters $k$, $\Pe$ and $\RR$ for several Gebhart numbers, together with the corresponding values of $\Pe$ and $\RR$ obtained by employing the $k \to 0$ asymptotic solution \equa{37}. The first row of this table, for $\Ge=0$, is relative to the asymptotic solution discussed in Section~\ref{smaGenum}. We note that the discrepancy between the asymptotic case $\Ge \to 0$ and the case $\Ge=0.01$ is less than $0.3\,\%$ both in terms of $k_c$ and $\RR_c$. This means that, for practical purposes, the asymptotic solution for $\Ge \to 0$ can be safely employed for all cases with $\Ge \le 0.01$. In Table~\ref{tab1}, we also reported the threshold value of $\Ge$ defined by \equa{36rees2} above which $k_c = 0$.

\begin{figure}[t]
\centering
\includegraphics[width=0.999\textwidth]{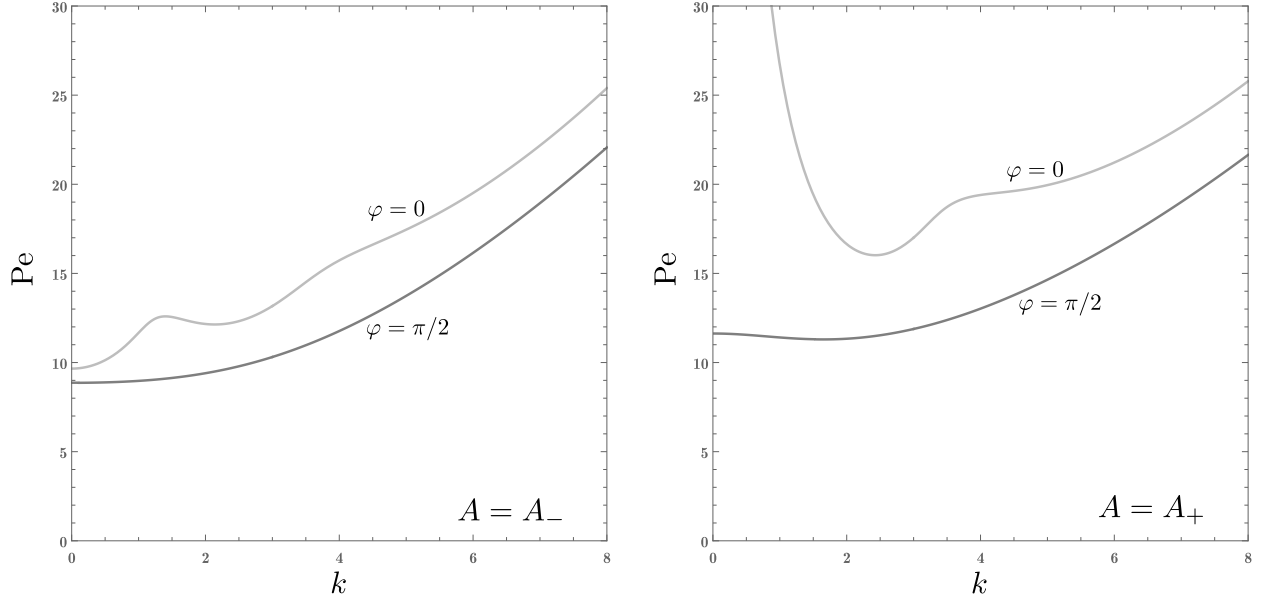}
\caption{\label{fig11}Neutral stability curves for $\Ge = 3.8$ in the $(k, \Pe)$ plane for transverse (light grey line, $\varphi=0$) and longitudinal (dark grey line, $\varphi=\pi/2$) modes. Both dual branches $A = A_{-}$ and $A = A_{+}$ are considered}
\end{figure}

Even if we declared that the focus of our stability analysis is on the $A = A_{-}$ branch, a sample comparison between the stability characteristics of dual flows for a given Gebhart number could be interesting. Thus, we have selected a case, $\Ge = 3.8$, very close to the maximum, $\Ge = \sqrt{15}$, so that the dual branches are not too different. We recall that the basic dual flows for $\Ge = 3.8$ are illustrated in Figs.~\ref{fig1} and \ref{fig3} and commented on in Section~\ref{badufl}. The neutral stability curves relative to both the dual branches with $\Ge = 3.8$ are displayed in Fig.~\ref{fig11} for transverse modes and for longitudinal modes. In both dual branches, the longitudinal modes are the most unstable. However, the  $A = A_{-}$ branch displays a critical value of $\Pe$ smaller than the critical value for the $A = A_{+}$ branch. Then, the $A = A_{-}$ is more unstable than the $A = A_{+}$ branch in this case. This outcome is not surprising as a similar behaviour was observed in a similar previous studies dealing with Darcy's flow in a porous channel \cite{barletta2009stability}. A minor aspect clearly visible in Fig.~\ref{fig11} is that the onset of instability with longitudinal modes, for the $A = A_{+}$ branch, happens with a nonzero wavenumber mode, On the other hand, $k_c = 0$ for the $A = A_{-}$ branch.

\section{Conclusions}
The onset of convective instability for the buoyant parallel flows in a horizontal plane channel with adiabatic walls has been studied. The action of buoyancy and the unstable behaviour are induced by the viscous dissipation for a flow with a given mass flow rate. 

The main features and, in particular, the duality of the basic parallel flows have been surveyed. The stability analysis has been focussed on the lower branch of the dual flows, as the higher branch turned out to display features utterly incompatible with the Oberbeck-Boussinesq approximation. 
The governing dimensionless parameters for the dual flows are the Gebhart number, $\Ge$, also known as the dissipation number, and the P\'eclet number, $\Pe$. 

The stability analysis has been formulated by evaluating the neutrally stable value of $\Pe$ for a given $\Ge$. Since the dual flows are mathematically defined only with $\Ge \le \sqrt{15}$, this whole parametric range has been explored. However, it has been also mentioned that values of $\Ge$ as large as its maximum are hardly significant for any real-world application, even over length scales of geophysical interest \citet{barletta2022mixed}. 

The linear stability analysis has been formulated by assuming creeping flow, which means an infinite Prandtl number. Such an analysis has been carried out through a numerical solution of the stability eigenvalue problem obtained by the shooting method. The main results of the stability analysis can be summarised as follows:
\begin{itemize}
\item The preferred perturbation modes causing the transition to instability are longitudinal, for every value of $\Ge$. The reaction of the base flows to arbitrary oblique perturbation modes has been tested by defining the inclination angle $\varphi$ between the base flow direction and the wave vector corresponding to the oblique modes.  Then, the inclination angle has been varied in the range $0 \le \varphi \le \pi/2$, with $\varphi=0$ identifying the transverse modes and $\varphi=\pi/2$ defining the longitudinal modes. The value $\varphi= \pi/2$ always yields the least stable condition.
\item The analysis of the neutral stability condition for the longitudinal modes reveals that such modes are non-travelling as their angular frequency and, hence, their phase velocity is always zero. The critical values of either $\Pe$ or the parameter $\RR = \Pe \sqrt{\Ge}$ are generally decreasing functions of $\Ge$, except for a very narrow range close to the maximum $\Ge = \sqrt{15}$. The longitudinal modes with a zero wavenumber, or an infinite wavelength, have a finite neutrally stable value of $\Pe$ for any nonzero $\Ge$. This regime has been studied via an analytical asymptotic solution.
\item A physically important situation is one where $\Ge$ is negligibly small, the P\'eclet number is very large, but the parameter $\RR$ remains finite. For this limiting case, the critical condition for the onset of the instability has been evaluated as $\RR_c = 41.8534$.
\end{itemize}

The study of the instability induced by the viscous dissipation has been carried out in this paper by assuming conditions of creeping flow where the Prandtl number is prescribed to be extremely large though maintaining a finite P\'eclet number. With this scenario in mind, the viscous dissipation instability is one emerging at very small Reynolds numbers and, hence, has likely no interrelation with the hydrodynamic instability analysed through the solution of the Orr-Sommerfeld problem. The extension of our study to a hybrid parametric domain where a finite Prandtl number is assumed can be a challenge for future investigations. Such a development can offer a chance to test the interplay between the viscous dissipation instability and the Orr-Sommerfeld hydrodynamic instability in a plane parallel channel.

\subsection*{Acknowledgement}
The authors acknowledge financial support from Italian Ministry of Education, University and Research (MIUR) grant number PRIN 2017F7KZWS.
%$$$$$$$$$$$$$$$$$$$$$$$$$$$$$$$

  \bibliographystyle{elsart-num-names}
  \bibliography{biblio}

%$$$$$$$$$$$$$$$$$$$$$$$$$$$$$$$ 

\end{document}